# Quantitative Study on Current-Induced Effect in an Antiferromagnet Insulator/Pt Bilayer Film

Pengxiang Zhang[1], Joseph Finley[1], Taqiyyah Safi[1], and Luqiao Liu[1]

[1]Department of Electrical Engineering and Computer Science, Massachusetts Institute of Technology, Cambridge, MA 02139, USA

**Abstract**

Quantitative investigation on the current-induced torque in antiferromagnets represents a great challenge, due to the lack of an independent method for controlling Néel vectors. Here by utilizing an antiferromagnetic insulator with Dzyaloshinskii-Moriya interaction, $\alpha$-$Fe_2O_3$, we show that the Néel vector can be controlled with a moderate external field, which is further utilized to calibrate the current-induced magnetic dynamics. We find that the current-induced magnetoresistance change in antiferromagnets can be complicated by resistive switching that does not have a magnetic origin. By excluding non-magnetic switching and comparing the current-induced dynamics with the field-induced one, we determine the nature and magnitude of current-induced effects in Pt/$\alpha$-$Fe_2O_3$ bilayer films.

Electrical control and detection of magnetic ordering inside antiferromagnets has attracted considerable interests, for potential advantages in operating speed and device densities. Current-induced magnetic switching has been recently reported in both metallic[1-7] and insulating[8-11] antiferromagnetic systems. In the former case, special crystal symmetries are utilized for realizing staggered spin-orbit torque, while in the latter one, the spin torque from an adjacent heavy metal layer is utilized for inducing magnetic dynamics. In these studies, anisotropic magnetoresistance (AMR), spin Hall magnetoresistance (SMR) or related planar Hall resistance (PHR) is generally employed to characterize the electrically induced 90° Néel vector switching. However, unlike ferromagnetic systems, where the current-induced spin torque is calibrated by using an external magnetic field as a standard[12-14], a quantitative relationship between the change of resistance value and the magnitude of spin torque in antiferromagnets remains to be built. So far X-ray based imaging techniques, which requires specialized facilities, have to be utilized to determine the ratio of switched magnetic domains[2,7,8,10]. Therefore, there is an urgent need for the development of an electrical measurement method that can be used to quantify the magnitude of current-induced effects in antiferromagnets.

$\alpha$-Fe$_2$O$_3$ is a well-studied antiferromagnetic insulator[15-17], with high Néel temperature (955 K) and strong antiferromagnetic exchange interaction (effective exchange field 900~1000 T). As is shown in Fig. 1(a), $\alpha$-Fe$_2$O$_3$ has trigonal crystal structure, and the two spin sublattices are stacked alternatively along (0001) direction. It is well-known[18] that at room temperature, $\alpha$-Fe$_2$O$_3$ exhibits an easy-plane anisotropy, which has a very weak ferromagnetism ($M_s$~2 emu/cm$^3$) due to the Dzyaloshinskii-Moriya (DM) interaction that causes a <0.1° in-plane canting angle between magnetic moments of the two sublattices. Because of the very weak magnetic anisotropy within the basal plane[18], the spin-flop field for aligning the Néel vector perpendicular to the external field direction in α-Fe$_2$O$_3$ can be very low (<1 Tesla), providing a convenient way for controlling the magnetic ordering orientation. In the meantime, the small net moment (**M**) is perpendicularly oriented with respect to the Néel vector (**N**) direction, allowing us to separate the different contributions from **M** and **N** to magnetic dynamics and transport effects such as SMR. Therefore, α-Fe$_2$O$_3$ represents a nice antiferromagnetic material platform, enabling us to characterize the electrically induced magnetic dynamics by comparing to field-induced ones.

We grew α-Fe$_2$O$_3$ films on α-Al$_2$O$_3$ (0001) substrates with magnetron sputtering and post-deposition annealing. As is shown in Fig. 1(b), despite the lattice mismatch of ~5.8%, epitaxial α-Fe$_2$O$_3$(0001) films were obtained. The magnetization hysteresis curves measured with a SQUID magnetometer at 300 K are illustrated in Fig. 1(c), which show a small $M_s$ around 1.5~2 emu/cm$^3$ and a coercive field of 500~1000 Oe within the entire thickness range of 10~120 nm, consistent with previously reported values[18,19]. In order to study current-induced switching and observe SMR, we sputter 5 nm Pt on α-Fe$_2$O$_3$ films and fabricate Hall bars of various widths [Fig. 1(d)], with the current and voltage channels aligned along [10$\bar{1}$0] and [$\bar{1}$2$\bar{1}$0] directions [Fig. 3(a)]. Both longitudinal ($R_{long}$) and transverse resistances ($R_H$) were measured with a rotating external field in the film plane, which aligns **N** perpendicular to the field orientation [Fig. 1(e) and (f)]. Consistent results have been obtained for $R_{long}$ and $R_H$ measurements after taking into account geometrical factors. In the following, we will mostly focus on $R_H$ measurement for monitoring magnetic dynamics. Close to saturation, $R_H$ curves can be fitted by $R_H = \frac{\Delta R_H}{2}\sin(2\theta_H)$, where $\theta_H$ is the field angle defined in Fig. 1(d). Our SMR signal is consistent with previous findings in other antiferromagnetic insulators such as NiO and Cr$_2$O$_3$[20-23]: $R_{long}$ reaches maximum when **N** is collinear with current. This is opposite to the SMR signal expected from a ferromagnet through the residual magnetic moment, suggesting the dominant role of Néel vector in the SMR effect. The magnitude of SMR ($\Delta R_{\text{long}}/R_{\text{long}} \sim 0.06\%$) is also comparable to earlier reports in NiO[20,21].

We first tested the current-induced 90° Néel vector switching by following the procedures in Ref. [8] and [10]. As is shown in Fig. 2(a), we send large current pulses with 10 ms pulse width that are oriented –45° (stage A) and +45° (stage B) from the horizontal direction to induce possible switching. According to the spin-orbit torque picture, this would flip *N* along the two orthogonal directions. To monitor the possible switching, we record $R_H$ after each pulse with a small sensing current of 2 mA. To minimize thermally induced resistance variation, $R_H$ is read after 10 s of waiting, allowing the device to return to equilibrium. As is shown in Fig. 2(b), a periodic change in $R_H$ is indeed observed. $R_H$ repeatedly switches between low and high values after applications of setting and resetting current pulses and forms a zigzag pattern.

The low spin-flop field in *α*-Fe$_2$O$_3$ allows us to control the Néel vector with a relatively small external field. Therefore, to check if the observed switching has magnetic origins, we compared the current-induced $R_H$ change with and without an external field. As is shown in Fig. 2(c), across the whole current range, the switching behaviors under these two conditions are very similar. The minor difference in the exact $R_H$ value between the two curves is smaller than run-to-run variations under the same field condition. From this comparison, we see that the applied field, which is supposed to align the Néel vector and suppress the current-induced switching, turns out to have negligible influence on the resistance change.

Furthermore, to understand the relationship between the resistance change and magnetic switching, we studied magnetic states after applied current pulses by measuring the angle-dependent SMR curves. In principle, if the switching behavior of Fig. 2(b) and (c) comes from the reorientation of Néel vector, the resistance value after these pulses would correspond to a peak or valley position in the subsequent angle-dependent SMR curve. However, we found no obvious change in SMR signal, which always starts from zero and oscillates identically as field rotates, regardless of the magnitude and direction of prior current pulses [Fig. 2(d) and (e)]. The main effect of the pulses is to cause an overall shift in SMR curve. Under large writing pulses, this shift can become even larger than the full range of field-induced SMR [Fig. 2(e)]. This independence between current-induced resistance switching and Néel vector reorientation suggests that within our studied current range, the observed switching has a pure resistive origin. Our results suggest that careful treatment is needed to distinguish switchings due to magnetic and non-magnetic origins. One possible explanation for our observed sawtooth-like switching could be the electromigration effect from Pt wire[24]. We estimate that the current-induced temperature change can be as high as ~150 K under the applied current[25], which is comparable to the values reported in previous memristive switching experiments with Pt[24].

The current-induced overall resistance shift represents an obstacle in revealing magnetic switching. To overcome this difficulty and search for possible features of magnetic dynamics, we designed a new experiment which allows the demonstration of magnetic switching tendency with smaller applied currents. Current-induced magnetic moment titling has been previously utilized in the 2$^{nd}$ harmonic configuration for quantitative determination of spin torque in ferromagnets[13,14] as well as for the qualitative check on Néel vector switching in an antiferromagnet[3]. Here we measured the angle-dependent SMR curve subject to an in-plane rotating field by applying different sensing currents. If there is a current-induced field or torque effect which tends to switch the magnetic ordering, the Néel vector will be tilted away (or towards) the current direction [Fig. 3(a)]. This will be reflected as a deviation from the original angle dependence of the SMR signal, from which we can extract the nature and magnitude of current's influence. Since we are focusing on the relative change of the SMR curve shape under constant currents, the overall shift due to resistive switching as in Fig. 2 does not make a contribution. Besides the current effect, we find that our α-Fe$_2$O$_3$ film always exhibits an constant, intrinsic easy-axis anisotropy within the basal plane due to the broken symmetry within the substrate, which favors [$\bar{1}2\bar{1}0$] as an easy axis for the Néel vector[25]. This easy-axis anisotropy energy is determined to be ~900 erg/cm$^3$ from our experiment. We note that our experimental approach below also applies to pristine α-Fe$_2$O$_3$ crystals with triaxial anisotropy within the basal plane[25].

The evolution of SMR signal with various currents applied along the +$x$ direction on a sample with 10 nm α-Fe$_2$O$_3$ and 10 μm channel width is shown in Fig. 3(b). The 2 mA curve is used as a baseline for its low current density (~4×10$^6$A/cm$^2$). As current increases, the field angle ($\theta_H$) corresponding to the peak and valley locations shifts towards the $x$ axis (0 and 180° direction), suggesting that a field that is closer to $x$ direction is needed to balance the current-induced effect for reaching the same SMR extrema. This observed trend, therefore, implies that the current's effect tends to align $M$ along $y$ axis (or equivalently $N$ along $x$ axis, since $M$ and $N$ are always perpendicular in this antiferromagnet). More quantitatively, the magnetic moment orientation $\theta_M$ can be determined from the SMR value together with the field scanning history information using the relationship of $R_H = \frac{\Delta R_H}{2}\sin(2\theta_M)$, where $\Delta R_H$ represents the peak to valley value under the lowest sensing current in Fig. 3(b). The current-induced Néel vector tilting can thus be characterized by the misalignment angle between $M$ and $H$ ($\theta_M - \theta_H$) as a function of $\theta_H$ [Fig. 3(c)]. We note that within the first half of the period ($\theta_H = 0$~90°), $\theta_M - \theta_H$ becomes less negative as current increases and at 20 mA it even switches sign, suggesting that current's effect is dominant over the intrinsic anisotropy under this current and $x$ axis becomes more energetically favorable for $N$. To examine the relationship between the current's effect and its flowing direction, we vary the current direction $\varphi$. We first set $\varphi = 90°$ by applying $I$ along $-y$. We find that the current's influence on magnetic anisotropy changes sign when compared with $\varphi = 0°$ case and $I$ now increases the intrinsic easy-axis anisotropy within the basal plane by tilting $N$ further towards the $y$ axis [Fig. 3(d)]. Next, we return to $\varphi = 0°$ but reverse the current direction, i.e., applying $I$ along $-x$ [Fig. 3(e)]. We find that the current-induced magnetic moment tilting shows similar trends under the reversal of current direction, i.e. both positive and negative currents induces an anisotropy along the $x$ axis. However, under large currents, a small difference appears between $\pm I$. Within the field angle range of $\theta_H = 0°{\sim}180°$, the positive current-induced tilting of Néel vector is larger, while it is smaller within $\theta_H = 180°{\sim}360°$. We also verify this current reversal effect by applying $\pm I$ under $\varphi = 90°$, where similar asymmetries were observed[25].

The current-induced magnetic moment titling shown in Fig. 3(b) to (e) can be summarized as a change in a net magnetic energy in $\alpha$-Fe$_2$O$_3$. Using its definition, we can extract this energy density[43] from the measured $\theta_M - \theta_H$ data through $E(\theta_M) = \mu_0 M H_{ext} \int_0^{\theta_M} \sin(\theta_H{'} - \theta_M{'}) d\theta_M{'}$, where $\mu_0$ is the vacuum permeability, $H_{ext}$ is the applied external field and $M$ is the net magnetization. $E(\theta_M)$ under a series of currents along the $+x$ ($-x$) direction are shown in Fig. 4(a) [Fig. 4(b)] for the same device. Based on the symmetry under current reversal, we can separate the current-induced energy change into two parts: the even component $\Delta E_{even} = [\Delta E(+I) + \Delta E(-I)]/2$ and the odd component $\Delta E_{odd} = [\Delta E(+I) - \Delta E(-I)]/2$. From Fig. 4, we see that $\Delta E_{even}$ follows $-\sin^2(\theta_M)$ under field rotation and has a 180° period, which represents a change in the uniaxial anisotropy, while $\Delta E_{odd}$ follows $\sin(\theta_M)$ and has a 360° period, which reflects a unidirectional magnetic energy (similar to a Zeeman energy). In the following, we will analyze the physics origin of these two effects.

First, to identify the origin of $\Delta E_{\text{even}}$, we carried out a systematical study on $\Delta E_{\text{even}}$ as a function of device width $w$ and $\alpha$-Fe$_2$O$_3$ thickness $t$, as is summarized in Fig. 4(c). $\Delta E_{\text{even}}$ shows a strong dependence on width $w$, but is independent of thickness $t$. Both characteristics are inconsistent with the spin-orbit torque mechanism as spin torque should only depend on $J$ regardless of the channel width, and inversely proportional to the magnetic layer thickness. The results in Fig. 4(c), however, agree with a thermal mechanism picture. Particularly the observed dependence on sample size is consistent with the Joule-heating-induced temperature increase, which has the form of $\Delta T \propto J^2 w$ [6,25,44]. While an overall temperature increase cannot explain the observed anisotropy energy change, the Joule heating can lead to an additional contribution through magnetoelastic coupling. Under Joule heating, the substrate lattice constant under the device region will increase due to thermal expansion, resulting in a net compressive stress [Fig. 4(a) inset]. Moreover, similar to other antiferromagnetic insulators, $\alpha$-Fe$_2$O$_3$ exhibits a fairly strong magnetoelastic effect with the reported[37,45] magnetostrictive coefficient $\lambda_s$ on the order of $10^{-6}$. Using a single parameter of $\lambda_s = 1.4 \times 10^{-6}$ as well as thermo-mechanical stress determined from finite element simulations[25], we find that the quadratic dependence of $\Delta E_{\text{even}}$ on $J$ obtained from devices with different $w$ and $t$, and under different current direction and magnitude[25], can all be well explained [see the solid curves in Fig. 4(c)].

We now turn to the odd component $\Delta E_{\text{odd}}$. In contrast to $\Delta E_{\text{even}}$, which shows a strong dependence on device size, $\Delta E_{\text{odd}}$ has the same linear dependence on $J$ across devices with all different $w$ [Fig. 4(d)], suggesting a non-thermal origin. As previously discussed, the $\sin(\theta_M)$ angle dependence of $\Delta E_{\text{odd}}$ in Fig. 4 (a) and (b) is consistent with the characteristic of a Zeeman energy. While either an Oersted field or a field-like torque $\tau_{\text{FL}}$ can contribute to this energy density, the comparison between the 5 and 10 nm thick $\alpha$-Fe$_2$O$_3$ samples excludes the Oersted field as the main mechanism. By comparing the two slopes in Fig. 4(d), we find that under the same current, $\Delta E_{odd}$ of the 5 nm sample is nearly twice the value of the 10 nm one, consistent with an interfacial mechanism like the spin-orbit torque. The small deviation from the expected factor of two might originate from the inaccuracy in thickness calibration. Moreover, from Fig. 4(d), we quantify the $\tau_{FL}$ efficiency in the 5 nm sample to be 25 Oe per $10^7 A/cm^2$, which is about 7 times larger than the Oersted field. The magnitude of $\tau_{FL}$ here, when normalized with the magnetic film thickness, is comparable to previously obtained values in ferromagnetic and ferrimagnetic insulators[13,26,46]. Finally, we note that unlike $\tau_{FL}$, which shows distinct symmetries from the current-induced magnetoelastic effect, the damping-like torque $\tau_{DL}$ induces magnetic moment tilting with the same $\sin^2(\theta_M)$ angular dependence as the magnetoelastic effect[25]. As $\tau_{DL}$ is only a function of $J$ and should not depend on $w$, we can estimate on the upper bound of $\tau_{DL}$'s contribution to the $\Delta E_{even}$ as ~ 150 erg/cm³ at $J = 1 \times 10^8 A/cm^2$ from Fig. 4(c), lower than the magnetoelastic effect with our smallest $w$. As shown in our simulation, for easy-plane antiferromagnets with large magnetostrictive coefficient such as $\alpha$-$Fe_2O_3$ and NiO[40,47] ($\lambda_s$ = 1~5 × $10^{-4}$), any influence from damping-like torque can be dominated by the magnetoelastic effect in the current experimental configuration.

To summarize, we experimentally studied current-induced magnetic dynamics in a heavy metal/antiferromagnetic insulator bilayer system. By calibrating current-induced effects with a magnetic field, we identified the two main contributions of current and determined their magnitude: the magnetoelastic effect and the field-like spin-orbit torque. The current-induced Néel vector tilting method enables the separation of real magnetic dynamics from non-magnetic resistive switching. Meanwhile, a systematic study on the device size dependence of current's effect allows to tell the spin-orbit torque and thermal effect apart. These approaches are applicable to other easy-plane antiferromagnets such as NiO where the spin-flop field remains relatively small (on the order of one Tesla) due to the weak anisotropy within the basal plane[25].

*Note Added*—During the preparation and revision of the manuscript, we became aware of related works of [48] and [49].

**Acknowledgements**

This work is supported in part by National Science foundation under award ECCS-1808826, and by SMART, one of seven centers of nCORE, a Semiconductor Research Corporation program, sponsored by National Institute of Standards and Technology (NIST). The material synthesis and characterization is partially supported by the National Science foundation under award DMR 14-19807 through the MRSEC shared facilities.

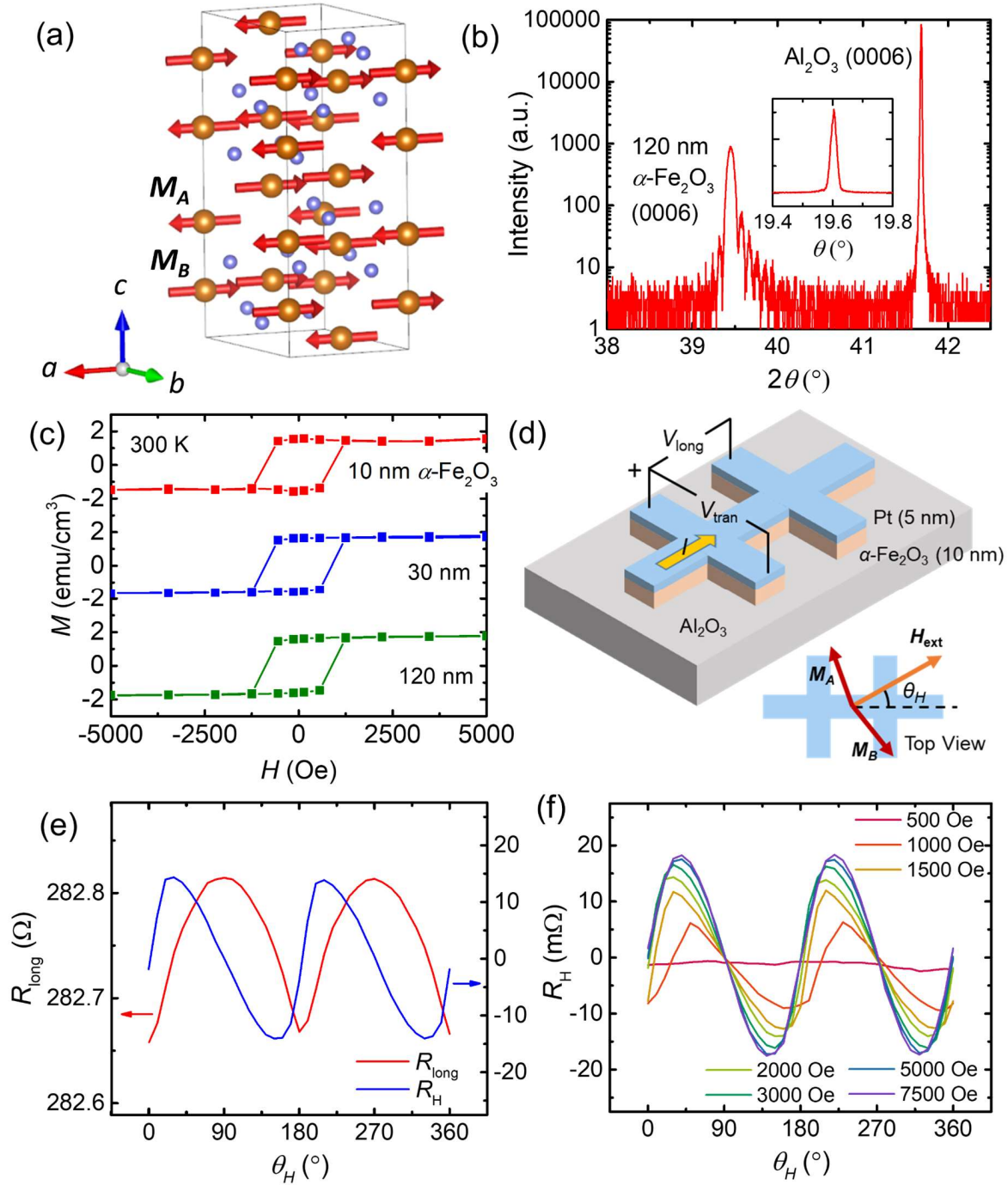

Figure 1 (a) Magnetic structure of α-$Fe_2O_3$. (b) X-ray diffraction of α-$Fe_2O_3$ (0001) film. Inset: rocking curve of the α-$Fe_2O_3$ (0006) peak. (c) SQUID magnetometry of α-$Fe_2O_3$ films for three different thicknesses. (d) Schematic of Hall bar device geometry and SMR measurement configuration. Inset: relative orientation between magnetic field and canted moments. (e) Longitudinal and transverse angle-dependent SMR under $H_{ext}$ = 2000 Oe. The sensing current is along [$\bar{1}2\bar{1}0$] direction. (f) $R_H$ under different in-plane fields.

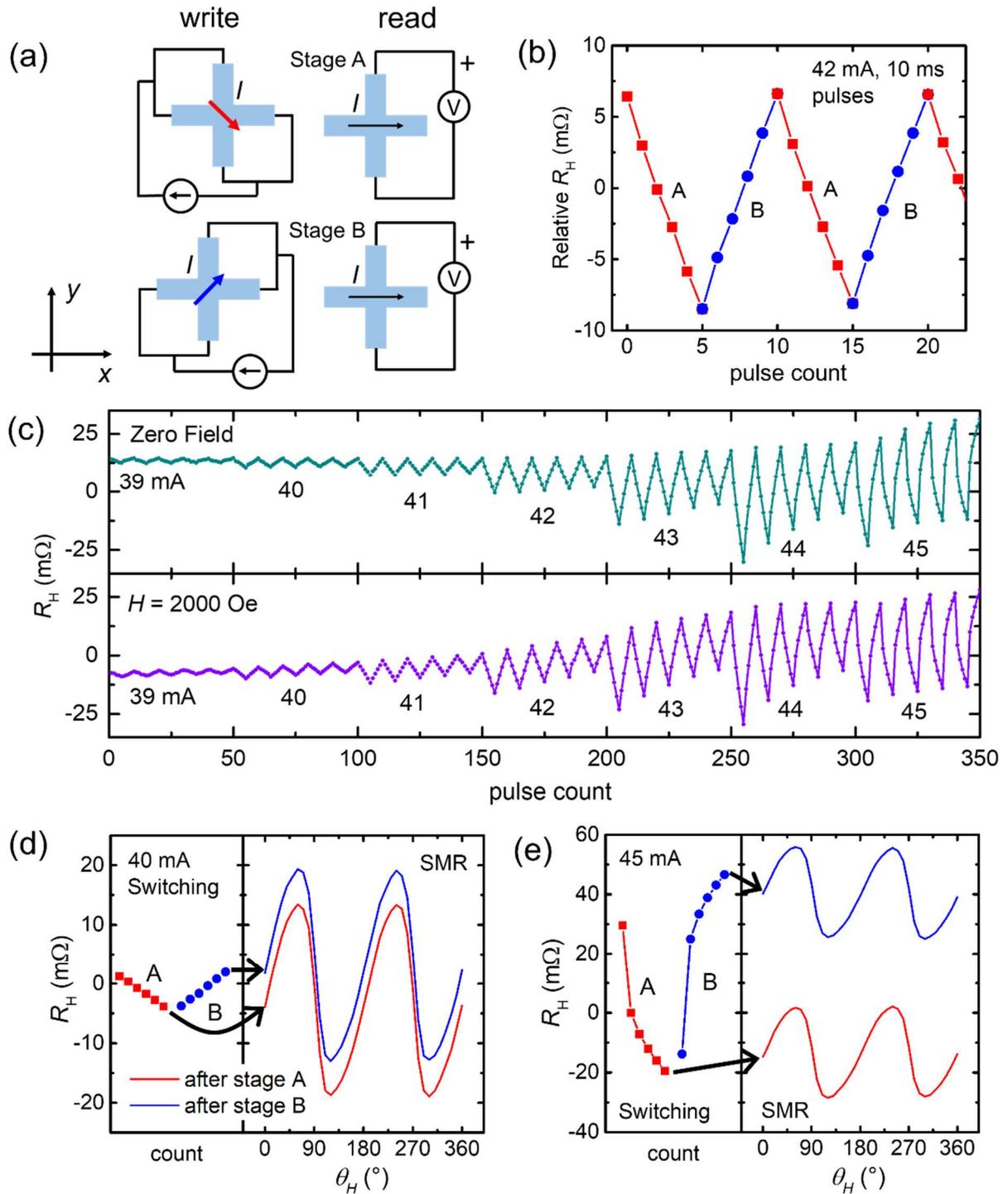

Figure 2 (a) Schematics of the writing and reading procedures. (b) Example of $R_H$ switching by current pulses. The red (blue) branches represent measured SMR signal in Stage A(B). The lateral dimension of the Hall cross is 10 μm × 60 μm, and the α-Fe$_2$O$_3$ thickness is 10 nm. (c)

Current-induced switching without and with $H_x = 2000$ Oe. (d) and (e) Measurement of angle-dependent SMR signal after writing current pulses of 40 mA and 45 mA, respectively, showing that the resistance change is not related to the SMR change.

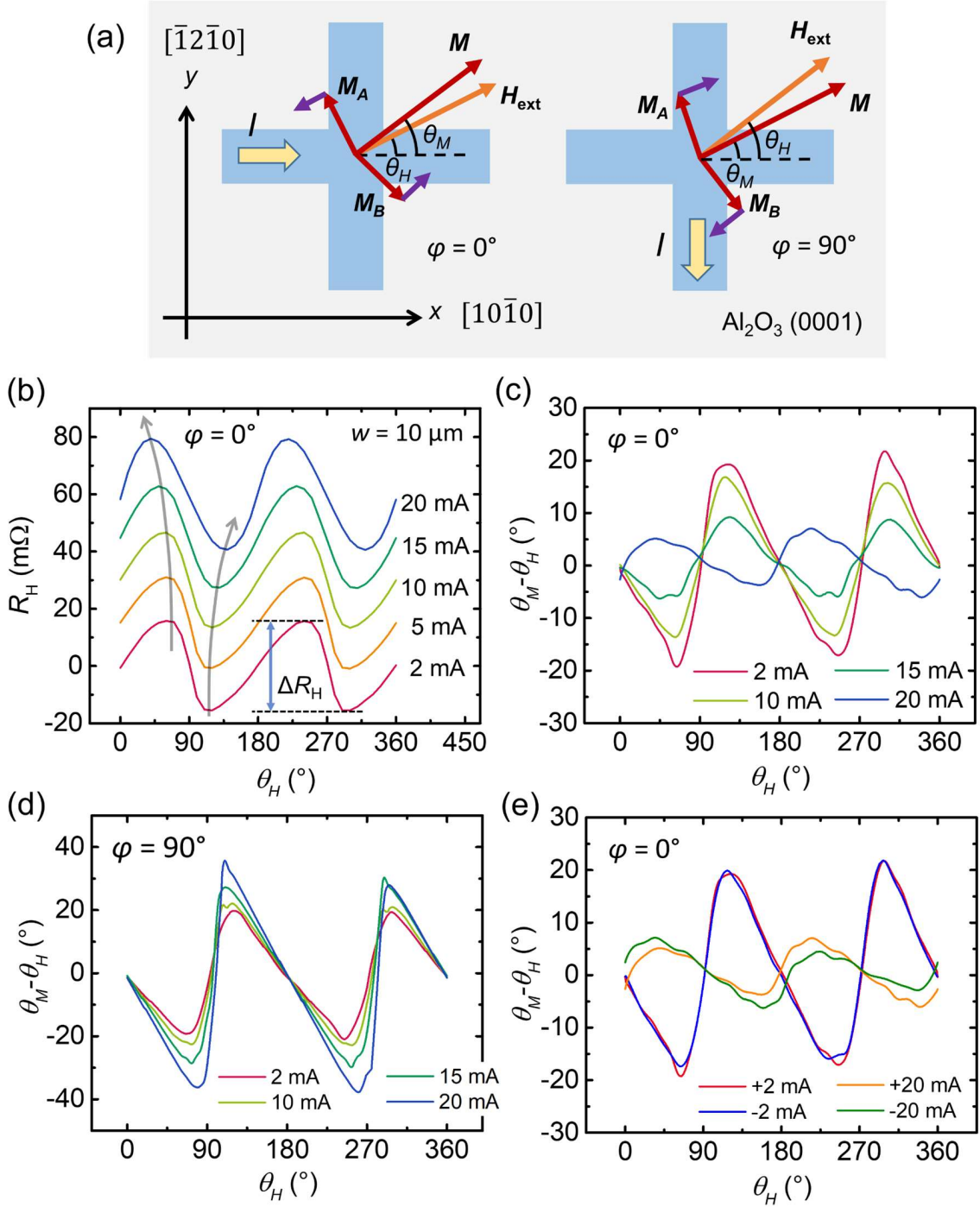

Figure 3 (a) Schematics of the magnetic moment tilting under applied currents. The left (right) panel is for current applied along $x(y)$ axis [$\varphi = 0°$ (90°)]. The purple arrows on sublattice moments indicate the tilting direction. Intrinsic magnetic anisotropy within basal plane is neglected. (b) Angle-dependent SMR under different currents for $\varphi = 0°$. The grey arrows are guide for the eye, illustrating the shift of peak and valley. $\Delta R_\text{H}$ is defined in the figure. (c) and (d) Angle between $\boldsymbol{M}$ and $\boldsymbol{H}$ as a function of $\theta_H$ for a range of applied currents at $\varphi = 0°$ and $90°$, respectively. The deviation of these curves from a perfectly smooth lineshape reflects the magnetic domain pinning effect from defect. (e) Comparison of current-induced magnetic moment tilting under positive and negative currents for $\varphi = 0°$, as is characterized by $\theta_M - \theta_H$. The results in Fig. 3 are obtained from a device with lateral dimension of 10 μm × 60 μm and α-$Fe_2O_3$ thickness of 10 nm.

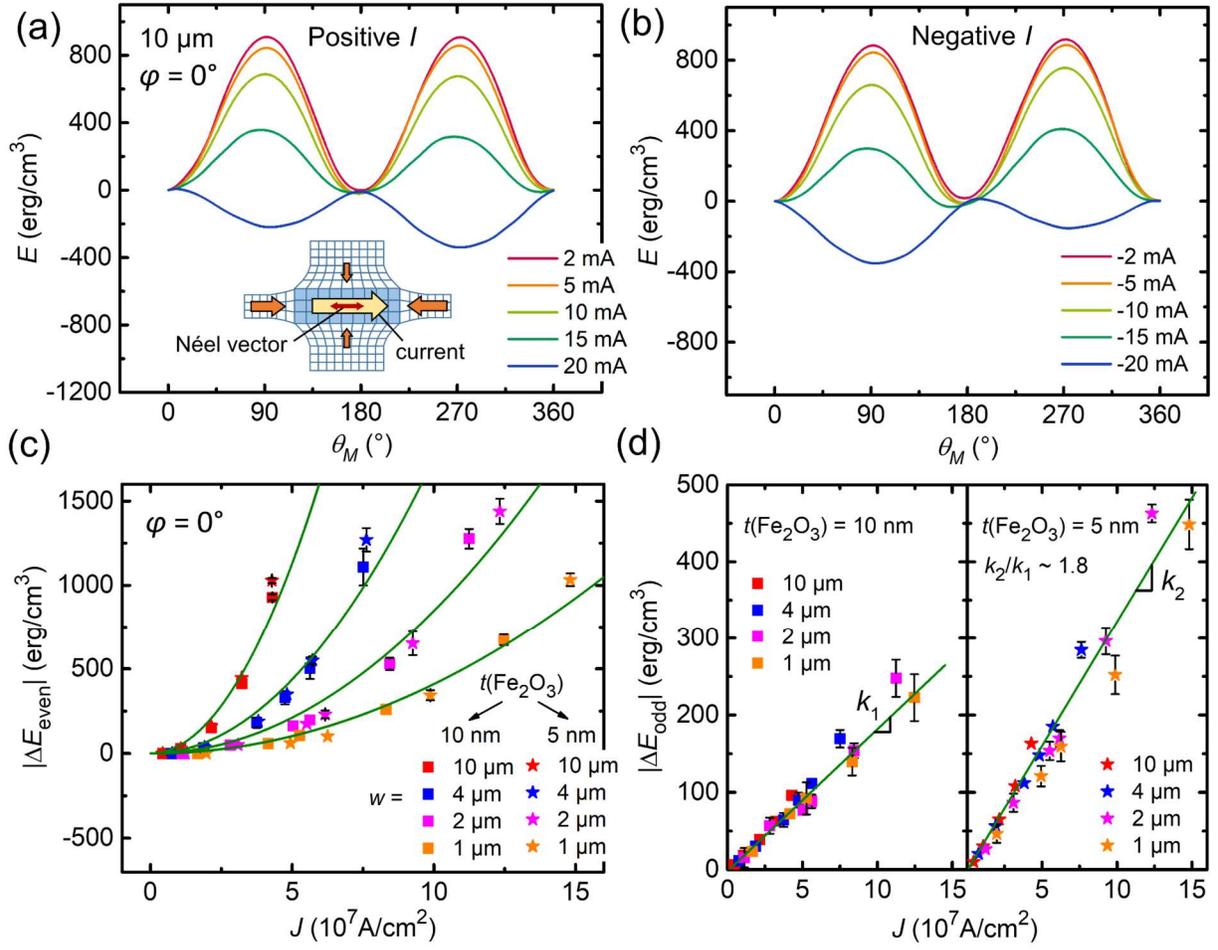

Figure 4 (a) and (b) Angle-dependent magnetic energy as a function of applied currents for positive (a) and negative (b) current polarities, on a 10 μm sample for $\varphi = 0°$. (a) inset: Schematic of Joule-heating-induced magnetoelastic effect. (c) and (d) The even and odd components of $\Delta E$ versus $J$ for devices with different $w$ and $\alpha$-$Fe_2O_3$ thickness $t(Fe_2O_3)$. The symbols and lines are results from experiment and calculation, respectively. The slopes for $t = 10$ nm and $t = 5$ nm lines in (d) have the ratio 1.8.